\documentclass[12pt]{iopart}
\usepackage{iopams}
\pdfoutput=1

\usepackage{graphicx}
 \newtheorem{theorem}{Theorem}

\begin{document}\today

\title{Iterative phase estimation}\author{C J O'Loan}\address{School of Mathematics and Statistics, University of St Andrews,  KY16~9SS, UK}\ead{cjo2@st-and.ac.uk}

\begin{abstract}
We give an iterative algorithm for phase estimation of a parameter $\theta$,
which is within a logarithmic factor of the Heisenberg limit. Unlike other methods,
we do not need any entanglement or an extra rotation gate which can perform arbitrary rotations with
almost perfect accuracy: only a single copy of the unitary channel and basic
measurements are needed. Simulations show that the algorithm is successful. We
also look at iterative phase estimation when depolarizing noise is present. It is
seen that the algorithm is still successful provided the number of iterative stages
is below a certain threshold.
\end{abstract}

\maketitle

\section{Introduction}
Phase estimation is of fundamental importance to quantum information and quantum computation. It is related to some very important problems such as estimating eigenvalues \cite{nori04,guzik05,wang08,wang09}, the factoring and search algorithms \cite[Section 5.3]{chuang00}, precision measurement of length and optical properties, and clock synchronization \cite{burghbart05}.

Suppose that we have a unitary matrix $U_\theta$ depending on an unknown parameter $\theta$ and that one of its eigenvectors $| u \rangle$ is completely known; 
furthermore $U_\theta$ acts on $| u \rangle$ in the following way:  $U_\theta| u \rangle = e^{i 2 \pi\theta} | u \rangle$, where $\theta \in [0,1)$.
The task of phase estimation is to estimate the eigenvalue $e^{i 2 \pi\theta}$, and consequently 
$\theta$, as accurately as possible.
In this paper we investigate phase estimation of a unitary matrix with known eigenvectors, which acts on a $2$-dimensional Hilbert space. In particular, we look at unitary matrices of the form
\begin{equation}
U_\theta = \left(  \begin{array} {cc}
1 & 0 \\
0 & e^{i 2 \pi \theta}  \ \end{array} \right),
\label{U}
\end{equation}
where $\theta \in [0,1)$.  We will think of $\theta$ as being a point on a circle of unit circumference, and confidence intervals for $\theta$ as arcs on a circle of unit circumference, known as confidence arcs.
We define the distance between an angle $\theta$ and an estimate $\hat \theta$ as
\begin{equation}
| \hat \theta - \theta |_{1} =\mathrm{min} \left( (\hat \theta - \theta )_{\mathrm{mod \, 1}}, (\theta - \hat \theta  )_{\mathrm{mod \, 1}} \right).
\end{equation}
It is important to recognise that the angle $\theta$ is an angle on the circle, and so in this paper all arithmetic is modulo $1$.

We now introduce the basic notions of quantum states and POVMs.
A quantum state is represented by a density matrix $\rho$, and satisfies $\tr \{ \rho \} =1$, $\rho = \rho^\dagger$, and $\rho \geq 0$.
In this paper we will mainly be dealing with pure states. Any state which can be written as $\rho = | \psi \rangle \langle \psi |$ is said to be a pure state; we will often refer to a pure state by its state vector $| \psi \rangle$.
Given a quantum state $\rho$, we can measure it using a POVM $M= \{ M_m \}$, obtaining outcome $m$ with probability given by the Born rule
\begin{equation}
p(m) = \tr \{ \rho M_m \}.
\end{equation}
A POVM $M=\{ M_m \}$ satisfies $M_m = M_m^\dagger$, $M_m \geq 0$ and $\sum_m M_m = \mathbb{I}$.

We shall quantify performance of phase estimation schemes in terms of the expected fidelity $\langle F( U_{\hat \theta},U_\theta) \rangle$ between $U_{\hat \theta}$ and $U_\theta$.
 We use the cost function
\begin{equation}
1- \langle F( U_{\hat \theta},U_\theta)\rangle = 1 - \frac{\left\langle|\tr \{ U_{\hat \theta}^{-1}  U_{\theta} \}|^2\right\rangle}{4}
\end{equation}
and look at its asymptotic scaling with $n$ -- the number of times that $U_\theta$ is used.

For a simple phase estimation approach where $U_\theta$ is used once on $n$ identical copies of some input state (see Section \ref{eq.simple}) we get $1- \langle F \rangle = O(1/n)$. This is known as the {\it standard quantum limit} \cite{burghbart05}.
However, it has been shown that when $n$ copies of the channel are available, \cite{hayashi06,kahn07,fujimai07}  we can get $1- \langle F \rangle = O(1/n^2)$. This rate at which $1- \langle F \rangle$ approaches zero is known as the {\it Heisenberg limit} \cite{gio04}, and cannot be beaten  \cite{kahn07}. 

In this paper we are interested in iterative phase estimation, when we have only a single copy of $U_\theta$, similar to that of Kitaev \cite{kit95}, for which   $1- \langle F \rangle = O((\log  n/n)^2)$.
This is within a logarithmic factor of the Heisenberg limit.
Note that there are other iterative phase estimation procedures with $1- \langle F \rangle = O(1/n^2)$, but they require an extra rotation gate capable of doing arbitrary rotations to high precision (see Section \ref{eq.bits}). 

In Section 2 a selection of different phase estimation schemes are described.
In Section 3 problems with some of the previous methods are explained. Section 4 contains our algorithm for phase estimation and a theoretical evaluation of its
performance. In Section 5 simulations are performed to check that our algorithm works. Section 6 looks at the performance
of our phase estimation algorithm in the presence of depolarizing noise.

\section{Phase estimation methods}

\subsection{Simple approach} \label{eq.simple}
A very simple method is to let $U_\theta$ act on the input state 
$| \psi_x \rangle = 1/\sqrt{2}(|0\rangle +|1\rangle)$; the output state is $| \psi_{\theta} \rangle = 1/\sqrt{2}(|0\rangle +e^{i2 \pi \theta}|1\rangle)$.
If we measure in $x$, that is we use the POVM $M_x = \{ M_1 = |\psi_x \rangle \langle \psi_x |, M_0 = \mathbb{I}-M_1 \} $, then we get outcome $1$ with probability $p_x(1; \theta) = (1+ \cos(2\pi \theta))/2$, and outcome $0$ with probability $p_x(0;\theta) = 1 - p_x(1; \theta)$.
If we perform $N$ measurements we get an estimate $ \cos(2 \pi \hat \theta) = 2 N_{x=1}/N -1$ of $\cos(2\pi\theta)$, where $N_{x=1}$ is the number of times we obtain the outcome $1$.  
If we measure in $y$, that is, use the POVM   $M_y = \{ M_1 = |\psi_y \rangle \langle \psi_y |, M_0 = \mathbb{I}-M_1 \} $, where $| \psi_y \rangle = 1/\sqrt{2}(|0\rangle + i|1\rangle)$,
we observe outcome $1$ with probability $p_y(1; \theta) = (1+ \sin(2\pi \theta))/2$, and outcome $0$ with probability $p_y(0;\theta) = 1 - p_y(1; \theta)$.
Performing $N$ measurements, we get an estimate  of $\sin(2\pi \theta)$;  from estimates of $\cos(2\pi \theta)$ and  $\sin(2\pi \theta)$ we are able to estimate $\theta$.

\subsection{Kitaev}
As far as we are aware, Kitaev \cite{kit95} gave the first $l$-stage iterative phase estimation procedure. (The number of stages $l$ is chosen before hand, and will depend on the precision desired and experimental limitations).
At the $k$th stage of Kitaev's procedure, $U_\theta$ acts $2^{k-1}$ times on a qubit, which is then measured.
We perform some multiple of $\log(l/\epsilon)$ measurements of $(2^{k-1}\theta)_{\mathrm{mod \, 1}}$. This ensures that we can ``localize each of the numbers $(2^{k-1}\theta)_{\mathrm{mod \, 1}}$ in one of the 8 intervals $[(s-1)/8,(s+1)/8] \, (s=0, \dots,7)$ with error probability $\leq \epsilon/l$.'' Using this information, an algorithm --- which is not given --- gives us an estimate $\hat \theta$ satisfying
\begin{equation}
\mathrm{Pr}\left(\left(\hat \theta -  \frac{1}{2^{l+2}}, \hat \theta +  \frac{1}{2^{l+2}} \right)  \ni \theta \right) \geq 1 - \epsilon.
\label{kitty}
\end{equation}

\subsection{Rudolph and Grover}
Rudolph and Grover \cite{rudolph03} looked at the problem of transmitting a reference frame from Alice to Bob, which is linked to estimation of  an unknown $U \in SU(2)$, parametrized by three parameters $\alpha, \theta, \phi$. The scheme of Rudolph and Grover involves estimating the parameters $\alpha, \theta, \phi$ individually using the following $l$-stage iterative procedure.
We set $\theta \in [0,1)$ to have an infinite binary expansion $\theta=w_1 w_2 \dots w_l \dots$.
At the $k$th stage a qubit is sent back and forth between Alice and Bob in such a way that, when Bob finally measures it, he obtains outcome $1$ with probability $p_k(1;\theta) = (1+ \cos(2^{k}\pi \theta))/2$.

This is repeated a minimum of $N = 32 \log_2 (2l / \epsilon)$  times \cite{rudolph03}, which ensures that Bob's estimate $\hat p_k(1;\theta)$ of $ p_k(1;\theta)$ satisfies
\begin{equation}
\mathrm{Pr}\left( \left( \hat  p_k - \frac{1}{4}, \hat p_k + \frac{1}{4} \right) \ni p_k \right) \geq 1 - \frac{\epsilon}{l}.
\label{RG270}
\end{equation}
It is assumed that if $|\hat p_k - p_k| \leq 1/4$, then Bob can estimate the $k$th bit of $\theta$ correctly.
If this is so, then from (\ref{RG270}), the probability that Bob estimates the $k$th bit of $\theta$ correctly is at least $1 - \epsilon/l$, and the probability that he estimates all of the binary digits of $\theta$ correctly is at least $1-\epsilon$. 
After $l$ stages, we get an estimate $\hat \theta =  \hat w_1 \hat w_2 \dots \hat w_l$,  satisfying
\begin{equation}
\mathrm{Pr}\left( \left( \hat  \theta - \frac{1}{2^{l}}, \hat \theta +\frac{1}{2^{l}} \right) \ni \theta \right) \geq 1 - \epsilon.
\end{equation}
A similar scheme is then used to estimate the parameters $\alpha$ and $\phi$. The method of Rudolph and Grover has been used in \cite{burghbart05} for the problem of clock synchronization. 

\subsection{Zhengfeng {\it et al}}
Zhengfeng {\it et al}  \cite{zheng06} highlighted two errors with the method of Rudolph and Grover:
(i) knowing $| \hat \theta - \theta|_1 \leq 1/2^m$ does not imply that we know the first $m$ bits of the binary expansion of $\theta$ ---  
consider $\theta = 0.49$, $\hat\theta=0.5$ and $m=1$, 
(ii) the method is problematic (in the sense explained in section 3) for $\theta$ close to $1/2$.

Zhengfeng {\it et al.}  gave the following $l$-stage procedure. In the first stage we let $U_\theta$ act on $| \psi_x \rangle$ and measure in $x$; we obtain outcome $1$ with probability $p(1;\theta) = (1+ \cos(2\pi\theta))/2$. The state  $U_\theta | \psi_x \rangle$ is measured some multiple of $\log(l/\epsilon)$ times ($N$), which gives an estimate $\hat \theta$ satisfying
\begin{equation}
\mathrm{Pr}\left( \left( \hat \theta - \frac{1}{12}, \hat \theta + \frac{1}{12} \right) \ni \theta \right) \geq 1 - \frac{\epsilon}{l}.
\label{1/12}
\end{equation}
Having obtained an estimate $\hat \theta$,
\begin{enumerate}
\item[1)] if $\hat \theta \in [0, 5/12)$, define $r_1=2$ and $\nu_1 =0$,
\item[  2)] if $\hat \theta \in [5/12, 7/12)$, define $r_1=3$ and $\nu_1 =1$,
\item[3)] if $\hat \theta \in [7/12, 1)$, define $r_1=2$ and $\nu_1 =1$.
\end{enumerate}
At the $k$th stage we apply $U_\theta$ $r_1 r_2 \dots r_{k-1}$ times. After measuring $U_\theta^{r_1 r_2 \dots r_{k-1}}|\psi_x \rangle$ $N$ times, we estimate $(r_1 r_2 \dots r_{k-1} \theta)_{\mathrm{mod \, 1}}$ and obtain $r_k$ and $\nu_k$ in a similar way to that in which  we obtained $r_1$ and $\nu_1$. 
After $l$ stages we have $(r_1, \dots,r_l, \nu_1, \dots, \nu_l)$.
Our final estimate of $\theta$ is
\begin{equation}
\hat \theta = \sum_{i=1}^l \frac{\nu_i}{\prod_{k=1}^i r_k}.
\label{eq.zhen1}
\end{equation}

\subsection{Dob\v{s}\'{\i}\v{c}ek {\it et al}} \label{eq.bits}

A popular iterative estimation method is to take $\theta$ to have a binary expansion of length $l$ plus some small remainder, that is $\theta = w_1 w_2 \dots w_l + \Delta$. The binary digits $w_1, \dots, w_l$ are measured one at a time with a single measurement.
This has been done in \cite{childs00,dobi06,knill07,higgins07}.
(Higgins {\it et al} \cite{higgins07} were the first to give, and carry out experimentally,
a method of this form which achieves the Heisenberg limit.)
We review the method as described by Dob\v{s}\'{\i}\v{c}ek  {\it et al} \cite{dobi06}. 

At the $k$th stage we let $U_\theta^{2^{l-k+1}}$ act on one of two qubits; the other qubit is acted on by a  $Z$-rotation gate $e^{i \alpha_k \sigma_z}$ before being measured ---  where  $\alpha_0 =0$ and $\alpha_k$ for $k=2, \dots, l$ depends on the results from the previous $k-1$ stages.
From this measurement we get an estimate $\hat w_{l+1-k}$  of the $(l+1-k)$th binary digit. After $l$ stages we get an estimate $\hat \theta = \hat w_1 \hat w_2 \dots \hat w_l$ of $\theta$ which satisfies
\begin{equation}
\mathrm{Pr}\left( \left(\hat \theta - \frac{1}{2^{l+1}} ,\hat \theta + \frac{1}{2^{l+1}} \right) \ni \theta \right) \geq  0.81.
\end{equation}

We can increase the probability that our final interval contains $\theta$ to $1-\epsilon$ by either (a) increasing the number of rounds to $l' = l + \log(2+1/(2\epsilon))$  or (b) using  $O(\log^2(1/\epsilon))$  extra measurements of the first few binary digits \cite{dobi06}. The method of
Dob\v{s}\'{\i}\v{c}ek   {\it et al} has recently been carried out on experimental data in \cite{liu07}.

The method of Dob\v{s}\'{\i}\v{c}ek {\it et al} \cite{dobi06} has been analysed when there are internal static imperfections and residual couplings between qubits \cite{gar08}. It was shown that this type of noise is detrimental to the performance of Dob\v{s}\'{\i}\v{c}ek's method, however,
solutions were found to overcome this problem in \cite{gar08}.

\section{Problems}
There is nothing wrong with Kitaev's method of iterative estimation. However, he does not give 
an algorithm for (i) choosing which of the intervals contains $(2^{k-1}\theta)_{\mathrm{mod \, 1}} $ with probability $1 - \epsilon/l$, (ii) reconstructing $\theta$ given confidence intervals for $(2^{k-1}\theta)_{\mathrm{mod \, 1}} $.
As we will see in this section there are gaps in the methods of Rudolph and Grover, and Zhengfeng {\it et al}  for (i).
 
There are two main gaps in Rudolph and Grover's method, which we now explain.
Firstly, $p_k(1;\theta) = (1+ \cos(2^{k}\pi \theta))/2$ is a multimodal function of $\theta$. For example, $\theta =3/4$ and $\theta'=1/4$ give the same value of $p_1(1;\theta)$, even though they differ in the first binary digit. To overcome this, we need an estimate of $\sin(2\pi \theta)$ as well.
This however is a trivial point and is easily overcome.

Secondly, if $\theta = 1/2 \pm \delta$, where $\delta$ is small, we require a large number of measurements to determine the first bit of $\theta$ correctly with high probability.
If we do make a mistake then for our final estimate $\hat \theta$ we will have $| \hat \theta - \theta|_1 \geq \delta$. 
This problem, which occurs for $\theta$ close to $1/2$, was pointed out by Zhengfeng {\it et al}.

A similar problem also occurs for $\theta= 0 \pm \delta$.
Because of this, we will encounter difficulties in estimating the $k$th bit of $\theta$ whenever
$(2^{k-1}\theta)_{\mathrm{mod \, 1}}  \approx 0$, $(2^{k-1}\theta)_{\mathrm{mod \, 1}}  \approx 1$ or $(2^{k-1}\theta)_{\mathrm{mod \, 1}}  \approx 1/2$. 
However, it may also be possible to overcome this issue using extra rotation gates in these cases.

There are also gaps with the method of Zhengfeng {\it et al}  \cite{zheng06}. 
Firstly, like Rudolph and Grover, they overlook the fact that $p_1(1;\theta) = (1+ \cos(2\pi\theta))/2$ is bimodal.
Secondly, the accuracy of their final estimate relies on the assumption that if $| \hat \theta - \theta|_1 \leq 1/12$ and $\hat \theta \in [0, 5/12)$, then $\theta \in [0, 1/2)$. This is not true, as we could have $\theta = -1/12 \not \in [0, 1/2)$.
Similarly, they assume that  if $| \hat \theta - \theta|_1 \leq 1/12$ and $\hat \theta \in [7/12, 1)$, then $ \theta \in [1/2, 1)$, which again is not true, as we could have $\theta = 1/12  \not \in [1/2, 1)$.
Again we will get problems at the $k$th stage if $(r_1 \cdots r_{k-1} \theta)_{\mathrm{mod \, 1}}    \approx 0$ or $(r_1 \cdots r_{k-1} \theta)_{\mathrm{mod \, 1}} \approx 1$.

\section{Our approach} 
This section contains our method for phase estimation. Firstly, we describe the method for going from 
confidence arcs for $\theta, (2\theta)_{\mathrm{mod \, 1}}, (4\theta)_{\mathrm{mod \, 1}}, \dots, (2^{l-1} \theta)_{\mathrm{mod \, 1}}$, of length $1/3$ and coverage probability at least $1-\epsilon/l$, to a
confidence arc for $\theta$ of length $1/(2^{l-1} \times 3)$ and  coverage probability at least $1-\epsilon$.
 Secondly, we describe how to get a confidence arc for $(2^{k-1} \theta)_{\mathrm{mod \, 1}}$, of length $1/3$ and coverage probability $1-\epsilon/l$.
 Thirdly, we use Hoeffding's inequality to calculate the number of measurements needed at each stage.
 Finally, we show that $ 1- \langle F(U_{\hat \theta}, U_\theta) \rangle = O((\log n/n)^2)$.

 \subsection{The iterative phase estimation algorithm} \label{32}
 In this section we introduce our method for computing a confidence arc for $\theta$ of length $1/(2^{l-1} \times 3)$ and  coverage probability $1-\epsilon$.
First we give an intuitive approach using examples.
For computational simplicity,  we look at confidence arcs of length $0.3$ and coverage probability $1$.
$L_k$ and $J_k$ will denote confidence arcs for $(2^{k-1}\theta)_{\mathrm{mod \, 1}}$ and $2^{k-1}\theta$ respectively, of length $0.3$ and coverage probability $1$. ( In our more general algorithm $L_k$ and $J_k$ will have length $1/3$ and coverage probability at least $1-\epsilon/l$.)
For the examples we choose $l = 3$.
\\
{\bf Example}
\\ 
Suppose that after doing some measurements of $U_\theta$, $U_\theta^2$ and $U_\theta^4$ we find
\begin{eqnarray}
L_1 &=& [0.6, 0.9 ]  \ni \theta \label{ex1l1} \\
L_2 &=& [0.3, 0.6 ] \ni (2\theta)_{\mathrm{mod \, 1}} \label{ex1l2} \\
L_3 &=& [0.8, 1.1 ] \ni (4\theta)_{\mathrm{mod \, 1}}. \label{ex1l3}
\end{eqnarray}
It follows from (\ref{ex1l1}) that 
\begin{equation}
2L_1= [1.2, 1.8] \ni 2\theta.
\label{2ex1l1}
\end{equation}
Using (\ref{ex1l2}) and  (\ref{2ex1l1}), it follows that
\begin{equation}
J_2 =  [1.3, 1.6]  \ni 2\theta.
\label{defj21}
\end{equation}
From (\ref{defj21}) we know that 
\begin{equation}
2J_2= [2.6, 3.2]  \ni 4\theta.
\label{2j21}
\end{equation}
Using (\ref{ex1l3}) and (\ref{2j21}) we get
\begin{equation}
J_3 = [2.8, 3.1]  \ni 4\theta.
\label{defj3}
\end{equation}

Using confidence arcs (\ref{ex1l1}), (\ref{ex1l2}) and (\ref{ex1l3}) for $\theta, (2\theta)_{\mathrm{mod \, 1}}$ and $ (4\theta)_{\mathrm{mod \, 1}}$ respectively, of length $0.3$ and coverage probability $1$,
we have derived a confidence arc (\ref{defj3}) for $4\theta$ of length $0.3$ and coverage probability $1$.
From this, we get a confidence arc for $\theta$ of length $0.3/2^{3-1}=0.075$ 
and coverage probability $1$, namely
\begin{equation*}
 (1/4)J_3 = [0.7, 0.775]  \ni \theta.
\label{defj3b}
\end{equation*}

Remember that we are looking at confidence arcs on a circle.
On the circle the arc $[1.2, 1.8]$ is equivalent to the arc $[0.2, 0.8]$, as is 
$[2.2, 2.8], [3.2, 3.8] \dots$.
Similarly $[2.6, 3.2]$ is equivalent to $[0.6, 1.2]$.

We define the symbol $\subset _1$ to mean that a confidence arc on the circle, is a subset of another confidence arc on the circle. Similarly, we define $\in_1$ to mean that a point is contained within an arc on the circle, e.g. $0.3 \in_1 [1.2, 1.8]$.
The previous example was rather simple in that 
$[0.3, 0.6] \subset_1 [1.2, 1.8]$ and $[0.8, 1.1] \subset_1 [2.6, 3.2]$.
 In general we cannot assume that $L_{k+1} \subset_1 2 J_{k}$.
 \\
 \\
{\bf General Algorithm}
\\ 
Our confidence arcs are now of length $1/3$ rather than $0.3$. Let us put
\begin{eqnarray*}
L_k &=& [x(k), x(k) +1/3], \quad x(k) \in [0,1), \\
J_k &=& [z(k), z(k)+1/3].
\end{eqnarray*}
As in the Example we use $2J_k$ and $L_{k+1}$ to find a confidence arc $J_{k+1}$. We insist that $J_{k+1} \subset 2 J_k$. For this we require that 
$z(k+1) \in [ 2z(k), 2z(k)+1/3]$.
 Assuming that $J_k \ni 2^{k-1} \theta$ and $L_{k+1} \ni (2^k \theta)_{\mathrm{mod \, 1}}$ then there are three possibilities.
For each possibility we give a figure, with a small vertical line representing the choice of the lower bound for $J_{k+1}$. (Note that $J_1 = L_1$.)

(i) The simplest possibility is that $L_{k+1} \subset_1 2J_k$. This occurs when $ ( x(k+1) - 2z(k) )_{\mathrm{mod \, 1}} \in [0,1/3)$.
We choose $J_{k+1}$ to have lower boundary $z(k+1) = 2z(k)+  (x(k+1) - 2z(k))_{\mathrm{mod \, 1}}$.

 \begin{figure}[htp]\centering\includegraphics[totalheight=0.09\textheight]{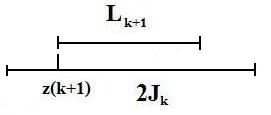}
 \caption[a]{Situation 1}
 \label{fig:conf1}\end{figure}

(ii) Another possibility is that $x(k+1) \not \in_1 2J_k$ but $x(k+1) +1/3 \in_1 2J_k$. 
This occurs when $ ( x(k+1) - 2z(k) )_{\mathrm{mod \, 1}} \in [2/3,1)$. In this case we take the
lower boundary of $J_{k+1}$ to be $z(k+1) = 2z(k)$.
 \begin{figure}[htp]\centering\includegraphics[totalheight=0.09\textheight]{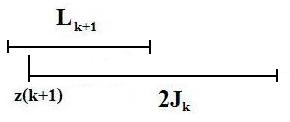}
 \caption[a]{Situation 2}
 \label{fig:conf2}\end{figure}

(iii) The final possibility is that $x(k+1) \in_1 2J_k$ but $x(k+1) +1/3 \not \in_1 2J_k$. 
This occurs when $( x(k+1) - 2z(k) )_{\mathrm{mod \, 1}} \in [1/3,2/3)$. In this case we take the
lower boundary of $J_{k+1}$ to be $z(k+1) = 2z(k)+1/3$.

 \begin{figure}[htp]\centering\includegraphics[totalheight=0.09\textheight]{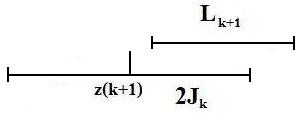}
 \caption[a]{Situation 3}
 \label{fig:conf3}\end{figure}

Using this iterative scheme, we end up with the confidence arc $J_l = [z(l), z(l)+1/3]$
for $2^{l-1}\theta$ with coverage probability $1$.
This immediately gives us a confidence arc for $\theta$ of length $1/(2^{l-1} \times 3)$,
namely $(1/2^{l-1}) J_l = [z(l)/2^{l-1}, (z(l)+1/3)/2^{l-1}]$.
We take the middle of this interval modulo $1$ as our estimate $\hat \theta$ of $\theta$, i.e. $((z(l)+1/6)/2^{l-1})_{\mathrm{mod \, 1}}$.

The final confidence arc for $\theta$ of length $1/(2^{l-1} \times 3)$ contains $\theta$ if 
$L_k \ni (2^{k-1}\theta)_{\mathrm{mod \, 1}}$, for every $k=1, \dots, l$.
If each $L_k$ has coverage probability at least $1-\epsilon/l$, the probability that every $L_k$ contains $(2^{k-1}\theta)_{\mathrm{mod \, 1}}$ is at least $1-\epsilon$. 

 \subsection{Finding $L_k$}
 Here we give the details of calculating the confidence arcs $L_k$ for $(2^{k-1} \theta)_{\mathrm{mod \, 1}}$ of length $1/3$ and coverage probability at least $1-\epsilon/l$.
  First we show how a confidence arc of length $1/3$ is computed,  then we show how to make the coverage probability at least $1-\epsilon/l$.
  We shall look at finding a confidence arc for $\theta$. The analysis is exactly the same as for $(2^{k-1} \theta)_{\mathrm{mod \, 1}}$ except in the latter case we let $U_\theta$ act $2^{k-1}$ times on the same $| \psi_x \rangle$.
 
 We let $U_\theta$ act on $| \psi_x \rangle$ and measure in $x$. We obtain outcome $1$ with probability
 $p_x(1;\theta) = (1+\cos(2\pi\theta))/2$.
We measure $U_\theta | \psi_x \rangle$ a total of $N$ times and obtain outcome $1$ $N_{x=1}$ times. We then have an estimate $2N_{x=1}/N -1$ of $\cos(2\pi\theta)$.

 We let $U_\theta$ act on $| \psi_x \rangle$ and measure in $y$. We obtain outcome $1$ with probability
 $p_y(1;\theta) = (1+\sin(2\pi\theta))/2$.
We measure $U_\theta | \psi_x \rangle$ in $y$ a total of $N$ times and obtain outcome $1$ $N_{y=1}$ times. We then have an estimate $2N_{y=1}/N -1$ of $\sin(2\pi\theta)$.
 We get an estimate 
 \begin{equation*}
 \hat \theta = \frac{1}{2\pi}\left(\mathrm{atan2}\left(\frac{2N_{y=1}}{N} -1, \frac{2N_{x=1}}{N} -1\right)\right)_{\mathrm{mod \, 2\pi}}
 \end{equation*}
 of $\theta$. We can construct $L_1$ as
 \begin{equation*}
 L_1 = \left( \left(\hat \theta -\frac{1}{6}\right)_{\mathrm{mod \, 1}}, \left(\hat \theta -\frac{1}{6}\right)_{\mathrm{mod \, 1}} +\frac{1}{3}\right).
 \end{equation*}
  More generally, given an estimate $(2^{k-1} \hat \theta_k)_{\mathrm{mod \, 1}}$ of $(2^{k-1} \theta)_{\mathrm{mod \, 1}}$ we get the confidence arc 
  \begin{eqnarray*}
 L_k &=& \left(x(k), x(k)+\frac{1}{3}\right),\\
 x(k) &=&\left(\left(2^{k-1} \hat \theta_k\right)_{\mathrm{mod \, 1}} -\frac{1}{6}\right)_{\mathrm{mod \, 1}}.
 \end{eqnarray*}
It is shown in the appendix that if
 \begin{equation}
 \left| \frac{N_{x=1}}{N} - p_x(1;\theta) \right| \leq 0.306  \label{hatpx}
\end{equation}
and
\begin{equation}
 \left| \frac{N_{y=1}}{N} - p_y(1;\theta) \right| \leq 0.306 , 
  \label{hatpy}
 \end{equation}
 then  $L_1 \ni \theta$.
 It follows that if
  \begin{equation}
\mathrm{Pr}  \left(\left| \frac{N_{x=1}}{N} - p_x(1;\theta) \right| \leq 0.306 \right) \geq \sqrt{1-\frac{\epsilon}{l}} \label{hatpx1}
\end{equation}
and
\begin{equation}
\mathrm{Pr}  \left(\left| \frac{N_{y=1}}{N} - p_y(1;\theta) \right| \leq 0.306 \right) \geq \sqrt{1-\frac{\epsilon}{l}} \label{hatpy1},
  \label{hatpy1}
 \end{equation}
 then
 \begin{equation}
 \mathrm{Pr} \left( L_1 \ni \theta \right) \geq 1 - \frac{\epsilon}{l}.
 \label{lkooh}
 \end{equation} 
An analogous result holds for $L_k, k=2, \dots,l$.
 It is shown below that if $N=5.34 \log (4l/\epsilon)$ then (\ref{hatpx1}) and (\ref{hatpy1}) hold.

\subsection{Number of measurements needed} \label{subsec:berny}
The Hoeffding inequality \cite{hoeff63} will be used.
\begin{theorem}
Given independent random variables $X_1,\dots, X_n$ with $a_i \leq X_i \leq b_i$,
then the following inequality holds for the sum $S_n = \sum_{i=1}^n X_i$:
\begin{equation}
\mathrm{Pr}\left( | S_n - E[S_n]| \geq nt \right) \leq 2 \exp\left(-\frac{2n^2t^2}{\sum_{i=1}^n (b_i - a_i)^2}\right).
\label{eq.Bern}
\end{equation}
\end{theorem}
The observed measurement outcomes from a single measurement in $x$ are independent random variables $X_i$ with $0 \leq X_i \leq 1$, and for which $S_n = N_{x=1}$ and $E[S_n] = np_x(1;\theta)$. 
Using (\ref{eq.Bern}) it is straightforward to show that
\begin{equation}
\mathrm{Pr} \left( \left| \frac{N_{x=1}}{N} - p_x(1;\theta) \right| \geq t  \right) \leq 2 \exp\left(-2Nt^2\right).
\label{eq.Bern2}
\end{equation}
From (\ref{eq.Bern2}) it can be shown that (\ref{hatpx1}) holds if 
 \begin{equation}
 N = 5.34 \ln \left(\frac{4l}{\epsilon}\right)
 \label{eq.Nx}
 \end{equation}
 measurements in $x$ are performed at each stage. The analysis is exactly the same for measurements in $y$,
 and so a total number of 
 \begin{equation}
 N_{tot} = 10.68 \ln \left(\frac{4l}{\epsilon}\right)
 \label{eq.ntot}
 \end{equation}
 measurements are required at each stage. This ensures that (\ref{hatpx1}) and (\ref{hatpy1}) hold, and consequently (\ref{lkooh}) holds.

\subsection{The behaviour of the fidelity} \label{534}
We now see how $1- \langle F(U_{\hat \theta}, U_\theta) \rangle$ scales with the number of times $U_\theta$ is used. 
As in \cite{rudolph03}, we look at the worst--case value of $1 - \langle F(U_{\hat \theta}, U_\theta) \rangle$. 
That is, if the final confidence arc does not contain $\theta$ then $\hat \theta = (\theta +1/2)_{\mathrm{mod} \, 1}$, and if it does then $\theta$ lies on the boundary of the confidence arc, i.e.\ 
 $|\hat \theta - \theta |_1 = 1/(2^{l} \times 3)$.
This gives
\begin{eqnarray*}
 1- \langle F(U_{\hat \theta}, U_\theta) \rangle &\leq& 1- \left((1 - \epsilon) \frac{1 + \cos(2\pi/(2^{l} \times 3))}{2} +  \epsilon \times 0\right) \\
&\approx& \epsilon +  \frac{\pi^2}{2^{2l}\times 9} - \frac{\epsilon \pi^2}{2^{2l}\times 9}.
\end{eqnarray*}
If we choose $\epsilon = 1/2^{2l}$, then $ 1- \langle F(U_{\hat \theta}, U_\theta) \rangle = O(1/2^{2l})$.
This requires a total of
\begin{equation}
N_{tot} = 10.68 \ln \left(4l \times 2^{2l}\right)
\label{eq.totn}
\end{equation}
measurements  at each stage.
The number of times $U_\theta$ is used is $n = N_{tot} (2^l-1)$,
and so $1/2^{l} \approx N_{tot}/n$. The number of measurements, (\ref{eq.totn}), made at each stage is $O(l)$;  noticing that $\log n$ is also $O(l)$, it follows that
\begin{equation}
 1- \langle F(U_{\hat \theta}, U_\theta) \rangle = O\left(\left(\frac{\log n}{n}\right)^2\right).
\label{eq.logn}
\end{equation}

\section{Simulations}
The analysis in Section \ref{534} concentrated on optimizing the worst-case asymptotic scaling of $1- \langle F \rangle$ with respect to $n$.
The cost function $1- \langle F \rangle$ is very
sensitive to outliers. A large number of measurements, (\ref{eq.totn}), were chosen so that the probability that the final interval did not cover $\theta$ was $1/2^{2l}$. This ensured that $1- \langle F \rangle$ was within a logarithmic factor of the Heisenberg limit. Rather than choosing $N_{tot}$
large to remove the large contribution of outliers, an experimenter
may be happy enough if the probability that $\theta$ is contained by his final confidence arc is greater than some value. This approach does not sacrifice precision, but rather an unnecessarily large coverage probability.

In this section we give a table of simulated results, for different numbers of
iterative stages, and different numbers of measurements at each stage. From this table we show how to calculate a confidence interval for the coverage probability.
An experimenter who wants a confidence arc for $\theta$ of certain length and coverage probability could look at the table and find the number of measurements needed to achieve this.

Simulations are performed with the computer package MAPLE. 
A value for the parameter $\theta \in[0,1)$ is given by a random variable with a uniform distribution.
Measurement results can be simulated, since the number of times outcome $1$ is observed has a Binomial distribution. For example, at the $k$th iterative stage, measuring in $x$,  $N_{x=1} \sim \mathrm{Bin}(N, (1+\cos(2^{k} \pi \theta))/2)$.
From the simulated results of measurements in $x$ and $y$ for stages $1, \dots, l$, an estimate of $\theta$ is obtained using the iterative algorithm given in Section \ref{32}. We can then test whether our final confidence arc contains $\theta$. This is done for $100,000$ randomly chosen $\theta$, and the number of times that $\theta$ is contained by the final confidence arc recorded.

For most recent iterative schemes the total number of iterations is reasonably small:
$6$ in \cite{higgins07} and $7$ in \cite{liu07}. We look at simulations with the number of iterations varying between $6$ and $9$. Table \ref{nni2b.tab} gives the number of times the final confidence arc contains the true value of $\theta$. 

\begin{table}[th]
\caption{Numbers of trials out of 100,000 with $| \hat \theta - \theta| \leq 1/(2^l \times 3)$.}
\bigskip
\centering
\begin{tabular}{ccccc}
\hline
&\multicolumn{4}{c}{Number of iterative stages $(l)$}\\
\cline{2-5}
$N_{tot}$    &  6 & 7 & 8 & 9  \\
\hline
20 &  99,792 & 99,729 & 99,747 & 99,712  \\
30  &  99,993 & 99,987 & 99,982 & 99,978  \\
40 &  99,999 & 100,000 & 99,998 & 99,999  \\
50  & 100,000 & 100,000 & 99,999  & 100,000 \\
\hline 

\end{tabular} 
 \label{nni2b.tab}
\end{table} 
 Using the above simulations the coverage probability can be estimated, i.e.\ the probability that, using the iterative algorithm, the known true value $\theta$ is contained in our final confidence interval.

Suppose the true (unknown) coverage probability is $p$.
For the $i$th trial put 
\begin{eqnarray*}
W_i  &=& 1 \quad  \mathrm{if \, interval \, covers}  \, \theta \\
  &=& 0 \quad \mathrm{if \, not.}
\end{eqnarray*}
Then $W_1, \dots, W_M$ are independent identically distributed Bernoulli random variables, i.e.\ $W_i \sim \mathrm{Bin}(1,p)$.
Thus
\begin{equation*}
W_1 +  \cdots + W_M \sim  \mathrm{Bin}(M,p).
\end{equation*}
If $m$ out of $M$ intervals cover $\theta$ then $p$ is estimated by $m/M$.
An approximate $95 \%$ confidence interval for $p$ is
\begin{equation*}
\frac{m}{M} \pm 1.96 \sqrt{\frac{m/M \left(1 - m/M \right)}{M}}. 
\end{equation*}
The longest confidence interval ($0.00066$) is that for using $9$ iterative stages and a total of $20$ measurements at each stage.
Using the half-length of this confidence interval, we can compute a confidence interval from the results given in Table \ref{nni2b.tab} with coverage probability at least $95\%$:
\begin{equation*}
\frac{m}{100,000} \pm 0.00033.
\end{equation*}
If an experimenter is content with a confidence arc of length no smaller than
$1/(2^9 \times 3) = 1/1536$ and estimated coverage probability no greater than $99.6\%$ then he need perform no more than 20 measurements at each stage. If the experimenter wanted to use even less measurements he could produce his own table of simulated results possibly even varying the number of measurements performed at each stage.

\section{The noisy case}
It is known that  when even a small amount of noise is present
the performance of phase estimation schemes is greatly reduced \cite{huelga97,shaji07}.  
This section investigates the performance of the iterative estimation algorithm when depolarizing noise is present. The channel
\begin{equation}
\rho_0 \mapsto (1-r) U_\theta \rho_0 U_\theta^\dagger + \frac{r}{2}\mathbb{I}_2, \qquad 0 < r < 1,
\label{eq.noisy}
\end{equation}
is considered, where $U_\theta$ is the same as before, (\ref{U}), and $\rho_0 = | \psi_x \rangle \langle \psi_x|$. (The channel (\ref{eq.noisy}) is identical to  $U_\theta \rho_0 U_\theta^\dagger$ undergoing phase damping with $\lambda = r(2-r)$  \cite[p.\ 383]{chuang00}.)
Ji {\it et al} \cite{zheng06} gave the very interesting result that if $r >0$,  then the optimal asymptotic rate at which $1- \langle F(U_{\hat \theta}, U_\theta) \rangle$ approaches zero is given by the standard quantum limit.

The whole point of using an iterative scheme is that the distinguishability of
$\theta$  from $\cos(n 2 \pi \theta)$, with $n >>1$, is considerably greater than from $\cos(2 \pi\theta)$. One measure of distinguishability is the Fisher information. Given a family of probability distributions with density functions $p(x; \theta)$, the Fisher information is defined as
\begin{eqnarray}
 F_\theta^M &\equiv&  \int   p(x;\theta) \left( \frac{\partial \ln p(x;\theta)}{\partial \theta} \right)^2 dx \label{eq.FI00a} \\
&=&  \int  \frac{1}{p(x;\theta)} \left( \frac{\partial p(x;\theta)}{\partial \theta} \right)^2 dx.
\label{eq:FI00}
\end{eqnarray}
Intuitively the Fisher information tells us the amount of `information' about a
parameter contained in a probability distribution. The Symmetric Logarithmic Derivative (SLD) quantum information
$H_\theta$ tells us the maximal attainable Fisher information obtained from measuring a state depending on an
unknown parameter \cite{braunsteincaves94}, i.e.
\begin{equation}
F_\theta^M \leq H_\theta.
\end{equation}
The SLD quantum information is defined in terms of the SLD quantum score $\lambda_\theta$ as
\begin{equation*}
H_\theta = \tr \{ \lambda_\theta \rho_\theta \lambda_\theta \},
\end{equation*}
where $\lambda_\theta$ is any self-adjoint solution of the matrix equation
\begin{equation*}
\frac{d \rho_\theta}{d\theta} = \frac{1}{2}(\rho_\theta \lambda_\theta +\lambda_\theta \rho_\theta).
\end{equation*}
To measure distinguishability, the quantity $F_\theta^M/m$ will be used, where $m$ is the number of times $U_\theta$ acts on the same input state.
This is because of interest is to maximize the distinguishability of $\theta$ per use of the channel.

If there is no noise, and the experimenter lets $U_\theta$ act $m$ times on the input state and measures in $x$, then outcome $1$ is observed with probability $p_x(1;\theta) = (1 + \cos (m 2 \pi \theta))/2$ and $0$ with probability  $p_x(0;\theta) = 1 - p_x(1;\theta)$. The Fisher information from this measurement is $F_\theta^{M_x} = 4 \pi^2 m^2$, which is equal to the SLD quantum information. Measuring in $y$ gives the same Fisher information. Thus $F_\theta^{M_x}/m = F_\theta^{M_y}/m = 4 \pi^2 m$.
At the $k$th stage of the iterative procedure, we let $U_\theta$ act $m=2^{k-1}$ times on the input state, and so $F_\theta^{M_x}/m = F_\theta^{M_y}/m = \pi^2 2^{k+1}$. Thus $F_\theta^{M}/m$ (where $M$ is an arbitrary measurement in $x$ or $y$) increases exponentially with $k$.

In the noisy case, letting $U_\theta$ act $m$ times on the output state and measuring in $x$, outcome $1$ is observed  with probability $p_x(1;\theta) = (1 + (1-r)^m \cos (m2 \pi \theta))/2$ and $0$ with probability  $p_x(0;\theta) = 1 - p_x(1;\theta)$. Measuring in $y$, outcome $1$ is observed with probability $p_y(1;\theta) = (1 + (1-r)^m \sin (m2 \pi \theta))/2$ and $0$ with probability  $p_y(0;\theta) = 1 - p_y(1;\theta)$.
This gives
\begin{eqnarray*}
F^{M_x}_\theta &=& \frac{4\pi^2m^2 (1-r)^{2m} \sin^2(2m \pi\theta)}{1-(1-r)^{2m} \cos^2(2m \pi\theta)}\\
 F^{M_y}_\theta &=& \frac{4\pi^2m^2 (1-r)^{2m} \cos^2(2m \pi\theta)}{1-(1-r)^{2m} \sin^2(2m \pi\theta)} \\
 H_\theta &=& 4\pi^2m^2 (1-r)^{2m}. 
\end{eqnarray*}
Notice that
\begin{equation*}
F^{M_x}_\theta + F^{M_y}_\theta \approx  H_\theta. 
\end{equation*}
Thus measuring both in $x$ and $y$, the average Fisher information from a single measurement $M$ is approximately $H_\theta/2$. 

 The maximal value of $F_\theta^M/m$, taken over $m$, will occur close to the maximal value of $H_\theta/m$.
When $r >0$, $H_\theta/m$, and hence $F_\theta^M/m$, does not increase indefinitely with $m$.  Instead it reaches its maximum at
\begin{equation}
m = - \frac{1}{2\log (1-r)},
\end{equation}
after which it decreases.
When $r$ is small, this maximum is obtained at 
\begin{equation}
m  \approx \frac{1}{2r}.
\label{bestrm}
\end{equation}
The number of stages that can be performed, for small $r$,  such that $H_\theta/m$, and hence $F_\theta^M/m$, increases at each stage is approximately $l \approx - \log_2 r$. A consequence of this
is that estimation close to the Heisenberg limit is not possible, asymptotically, when there is any depolarizing noise. This gives an alternative insight into the result of Ji {\it et al} \cite{zheng06}.

\begin{figure}[htp]
 \centering\includegraphics[totalheight=0.42\textheight]{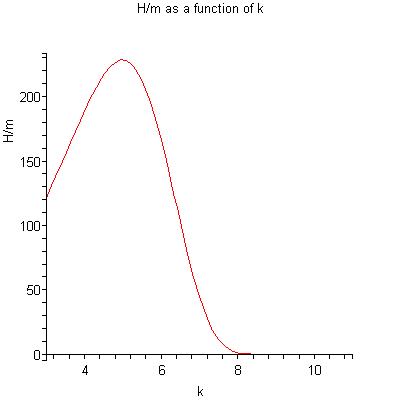}
 \caption[a]{$H_\theta/m$ at the $k$th iterative stage, with $r=2^{-5}$.}
 \label{fig:P5}
 \end{figure}

Figure \ref{fig:P5}  gives $H_\theta/m$ at the $k$th iterative stage when $r=2^{-5}$. It can be seen that $H_\theta/m$ increases up to $k = 5$, decreases slightly near $k = 6$ and falls rapidly for $k > 6$. Other figures not included here give similar information, showing $H_\theta/m$ increasing up to $k = - \log_2 r$, and decreasing rapidly for $k > - \log_2 r$.

Table \ref{nni30.tab} contains the results of simulations, for  magnitudes of noise $r=2^{-4},2^{-5}, \dots, 2^{-8}$ and total number of iterative stages $l = 4, \dots,9$.
Consider the diagonal of Table \ref{nni30.tab}, from $r = 2^{-4}$, $l =4$ to $r = 2^{-8}$, $l =8$. This corresponds 
to the experimenter performing $l = -\log_2 r$ iterative stages, which involves going up to the iterative stage at which $F_\theta^M/m$ is maximized. 
Similarly, the diagonal from $r = 2^{-4}$, $l =5$ to $r = 2^{-8}$, $l =9$  corresponds 
to the experimenter performing $l = -\log_2 r + 1$ iterative stages etc.
It is interesting to note that when $l >  -\log_2 r$, there is a significant decrease in the number of  confidence intervals containing $\theta$. If the experimenter performs $l = -\log_2 r$ iterative stages then the final confidence interval contains $\theta$ approximately $98\%$ of the time; for $l = -\log_2 r +1$ iterative stages, the coverage probability decreases to approximately  $89\%$. For $l = -\log_2 r +2$ iterative stages, the coverage probability is approximately $61\%$ -- a considerable drop in performance. Simulations using more measurements at each stage have given similar results.

It is interesting to see that the drop off in performance, in terms of the coverage probability,  occurs at the same point as the drop in performance as measured by $H_\theta/m$, and consequently $F_\theta/m$ -- seen in Figure \ref{fig:P5}.

We suggest, more generally, that for the channel (\ref{eq.noisy}) the optimum number of iterative stages, where at the $k$th stage $U_\theta$ is used $2^{k-1}$ times,  is $l = \lfloor-\log_2 r  \rfloor$. 

A related question was considered in \cite{rubin07}, where the `stopping point', was $N$ the number of entangled photons to be included in the NOON input states. Rubin and Kaushik found that the optimal precision in measurement occurred for $N=1.279/L$, where $L$ is the magnitude of loss (analogous to the point, $n'=1/(2r)$, at which $F_\theta^M/m$ is maximized).  

\begin{table}[th]
\caption{Numbers of trials out of 100,000 with $| \hat \theta - \theta|_1 \leq 1/(2^l \times 3)$,
with $N_{tot} =30$.}
\bigskip
\centering
\begin{tabular}{ccccccc}
\hline
&\multicolumn{6}{c}{Number of iterative stages $(l)$}\\
\cline{2-7}
$r$ &   4 & 5 & 6 & 7 & 8 & 9  \\
\hline
$2^{-4}$ &  98,290 & 88,340 & 60,423 & 32,445 & 16,059 & 8,042  \\
$2^{-5}$ &  99,804 & 98,408 &  88,537 & 61,293 & 32,756 & 16,460  \\
$2^{-6}$ &   99,967 & 99,807 &  98,430 & 88,708 & 61,148 & 32,595  \\
$2^{-7}$ &   99,985 & 99,955 &  99,802 & 98,476 & 88,895 & 61,699  \\
$2^{-8}$  &  99,988 & 99,977 &  99,962 & 99,812 & 98,467 & 88,864  \\
\hline 

\end{tabular} 
 \label{nni30.tab}
\end{table} 

\section{Discussion}
After completing this work we were made aware of similar work already done by
Higgins {\it et al} \cite{higgins09}.
In \cite{higgins09} it was shown that the logarithmic factor can be removed, thus achieving the Heisenberg limit. Also, an experimental demonstration was given.
However, an explicit algorithm which allows experimenters to implement this method was not given. Furthermore, the problem of noise was not dealt with.

\section{Conclusion}
In this paper we have shown that there are gaps in the iterative phase estimation schemes of \cite{rudolph03,zheng06}.

We have shown how to compute confidence arcs for $\theta, (2\theta)_{\mathrm{mod \, 1}}, (4\theta)_{\mathrm{mod \, 1}}, \dots, (2^{l-1} \theta)_{\mathrm{mod \, 1}}$, of length $1/3$ and coverage probability at least $1-\epsilon/l$.
The main contribution of this paper has been to give an explicit algorithm which uses these confidence arcs to obtain a confidence arc for $\theta$ of length $1/(2^{l-1} \times 3)$ and  coverage probability $1-\epsilon$.
Choosing $\epsilon = 1/2^{2l}$ gives $ 1- \langle F(U_{\hat \theta}, U_\theta) \rangle = O((\log n/n)^2)$, i.e. within a logarithmic factor of the Heisenberg limit.
The advantage of our scheme is that unlike other iterative phase estimation methods, such as \cite{dobi06,higgins07}, it does not require an extra rotation gate capable of doing arbitrary rotations with almost perfect accuracy. Thus our scheme has a simpler experimental setup and less potential for error. 

Using computer simulations we have shown that the algorithm is successful.
We have suggested the use of tables of simulated results to help choose the number of resources needed for desired levels of precision and coverage probability.

We have analysed our estimation scheme in the presence of depolarizing noise with magnitude $r$. We have shown that the iterative algorithm is still successful in this case provided that no more than $l=-\log_2 r$ iterative stages are performed.

\ack
Thanks go to Peter Jupp for his supervision and many helpful comments. This work was supported by an EPSRC Doctoral training grant. Thanks also to the referees for helpful suggestions, including result (\ref{referee1}).

\appendix
\section*{Appendix}
\setcounter{section}{1}

Put $x = \cos(2\pi \theta)$,  $y = \sin(2\pi \theta)$, $x_0 = 2N_{x=1}/N-1$, $y_0 = 2N_{y=1}/N-1$, $\phi(x,y) = \mathrm{atan2}(y,x)$ and $\hat \phi(x_0,y_0) = \mathrm{atan2}(y_0,x_0)$. Define 
\begin{equation*}
\Delta \hat\phi = \mathrm{min}  \left( (\hat \phi - \phi )_{\mathrm{mod \, 2\pi}}, (\phi - \hat \phi  )_{\mathrm{mod \, 2\pi}} \right). 
\end{equation*}
Given $\alpha \in [0, 1/\sqrt{2}]$ and 
\begin{eqnarray} 
| x- x_0| &\leq&  \alpha,  \label{xalpha1} \\
| y- y_0| &\leq&  \alpha, \label{yalpha1}
\end{eqnarray}
then $(x_0,y_0)$ lies in a square with sides of length $2\alpha$ centred around $(x,y)$. From simple geometry it is obvious that the maximum value of $\Delta \hat\phi$ occurs when $(x_0,y_0)$ is one of the four corners of the square. For this case, consider the triangle given by the points $(0,0)$, $(x,y)$ and $(x_0,y_0)$. The angle at point $(0,0)$ is $\Delta \hat\phi$, and  is opposite a side of length $\sqrt{2}\alpha$. The angle at point $(x_0,y_0)$ is opposite a side of length $1$.
It follows from the sine rule, and monoticity of $\arcsin$ on $[0,1]$ that
\begin{equation}
\Delta \hat\phi \leq \arcsin\left(\sqrt{2} \alpha\right).
\label{referee1}
\end{equation}
For the iterative algorithm it is required that $\Delta \hat\phi \leq \pi/3$, which holds if $\alpha = 0.612$. Then (\ref{xalpha1}) and (\ref{yalpha1}) are equivalent to (\ref{hatpx}) and (\ref{hatpy}).

\section*{References}
\bibliography{Biblio}

\end{document}